\documentclass[fleqn,10pt]{wlscirep}
\title{Tomonaga-Luttinger spin liquid in the spin-1/2 inequilateral diamond-chain compound K$_3$Cu$_3$AlO$_2$(SO$_4$)$_4$}

\author[1,*]{Masayoshi Fujihala}
\author[1]{Hiroko Koorikawa}
\author[1]{Setsuo Mitsuda}
\author[2]{Katsuhiro Morita}
\author[2]{Takami Tohyama}
\author[3]{Keisuke Tomiyasu}
\author[4]{Akihiro Koda}
\author[4]{Hirotaka Okabe}
\author[5]{Shinichi Itoh}
\author[5]{Tetsuya Yokoo}
\author[5]{Soshi Ibuka}
\author[6]{Makoto Tadokoro}
\author[6]{Masaki Itoh}
\author[7]{Hajime Sagayama}
\author[7]{Reiji Kumai}
\author[7]{Youichi Murakami}

\affil[1]{Tokyo University of Science, Department of Physics, Tokyo, 162-8601, Japan}
\affil[2]{Tokyo University of Science, Department of Applied Physics, Tokyo, 125-8585, Japan}
\affil[3]{Tohoku University, Department of Physics, Sendai, 980-8578, Japan}
\affil[4]{High Energy Accelerator Research Organization, Muon Science Laboratory and Condensed Matter Research Center, Institute of Materials Structure Science, Tsukuba, 305-0801, Japan}
\affil[5]{High Energy Accelerator Research Organization, Neutron Science Division, Institute of Materials Structure Science, Tsukuba, 305-0801, Japan}
\affil[6]{Tokyo University of Science, Department of Chemistry, Tokyo, 162-8601, Japan}
\affil[7]{High Energy Accelerator Research Organization, Photon Factory, Institute of Materials Structure Science, Tsukuba, 305-0801, Japan}

\affil[*]{fujihara@nsmsmac4.ph.kagu.tus.ac.jp}

\begin{abstract}
K$_3$Cu$_3$AlO$_2$(SO$_4$)$_4$ is a highly one-dimensional spin-1/2 inequilateral diamond-chain antiferromagnet. 
Spinon continuum and spin-singlet dimer excitations are observed in the inelastic neutron scattering spectra, which is in excellent agreement with a theoretical prediction: a dimer-monomer composite structure, where the dimer is caused by strong antiferromagnetic (AFM) coupling and the monomer forms an almost isolated quantum AFM chain controlling low-energy excitations.
Moreover, muon spin rotation/relaxation spectroscopy shows no long-range ordering down to 90~mK, which is roughly three orders of magnitude lower than the exchange interaction of the quantum AFM chain.
K$_3$Cu$_3$AlO$_2$(SO$_4$)$_4$ is, thus, regarded as a compound that exhibits a Tomonaga-Luttinger spin liquid behavior at low temperatures close to the ground state.
\end{abstract}
\begin{document}

\flushbottom
\maketitle
\thispagestyle{empty}

\section*{Introduction}
Identifying spin liquid phases in the ground state is one of hot topics in the field of low-dimensional quantum magnets. Three-dimensional orders induced by magnetic interactions between one-dimensional (1D) chains/two-dimensional layers, however, prevent the spin-liquid ground state. It is thus essential to search for quantum magnets that possess negligibly weak inter chain/layer interactions. 

One of possible spin liquids in 1D is a Tomonaga-Luttinger (TL) liquid, where spin-spin correlation decays algebraically with distance.
Azurite Cu$_3$(CO$_3$)$_2$(OH)$_2$ that contains spin-1/2 distorted diamond chains~\cite{Kikuchi,Rule_2005,Rule_2008,Rule_2011} is a possible candidate for the TL spin liquid, since the ground state of the distorted diamond-chain is expected to belong to an alternating dimer-monomer phase where neighboring  monomers are connected via the dimer in between and an effective Heisenberg 1D chain controls low-energy excitations \cite{Okamoto_1999,Okamoto_2003}.
In fact, a recent theoretical approach based on density functional theory together with numerical many-body calculations has proposed a microscopic model for the azurite, which includes two energy scales coming from dimer singlet and from a 1D Heisenberg chain \cite{Jeschke}.  Although the model predicts the TL spin liquid, three-dimensional magnetic interactions in the azurite cause the magnetic order at 1.85~K. 

Recently a highly 1D inequilateral diamond-chain compound alumoklyuchevskite K$_3$Cu$_3$AlO$_2$(SO$_4$)$_4$ has been reported by some of the present authors\cite{Fujihala}. There is no long-range magnetic order down to at least 0.5~K as evidenced by specific heat measurements. The magnetic susceptibility exhibits a double broad peak at around 200~K and 50~K. By analyzing the magnetic susceptibility, an effective model for K$_3$Cu$_3$AlO$_2$(SO$_4$)$_4$ has been proposed~\cite{Morita}, where the dimer is formed by one of the four sides in the diamond and the remaining spins form a 1D Heisenberg chain, as shown in Fig.~\ref{config}(d). The model is different from that for the azurite. 

In this paper, we present detailed studies of K$_3$Cu$_3$AlO$_2$(SO$_4$)$_4$ through single-crystal and powder X-ray diffraction (XRD), inelastic neutron scattering (INS), and muon spin rotation/relaxation ($\mu$SR) spectroscopy. 
These experimental results indicate that K$_3$Cu$_3$AlO$_2$(SO$_4$)$_4$ is an appropriate model material for the investigation of the TL spin liquid state at low temperatures close to the ground state.

\section*{Results and Discussion}

\subsection*{Crystal structure refinement}
\begin{figure}[ht]
\centering
\includegraphics[width=\linewidth]{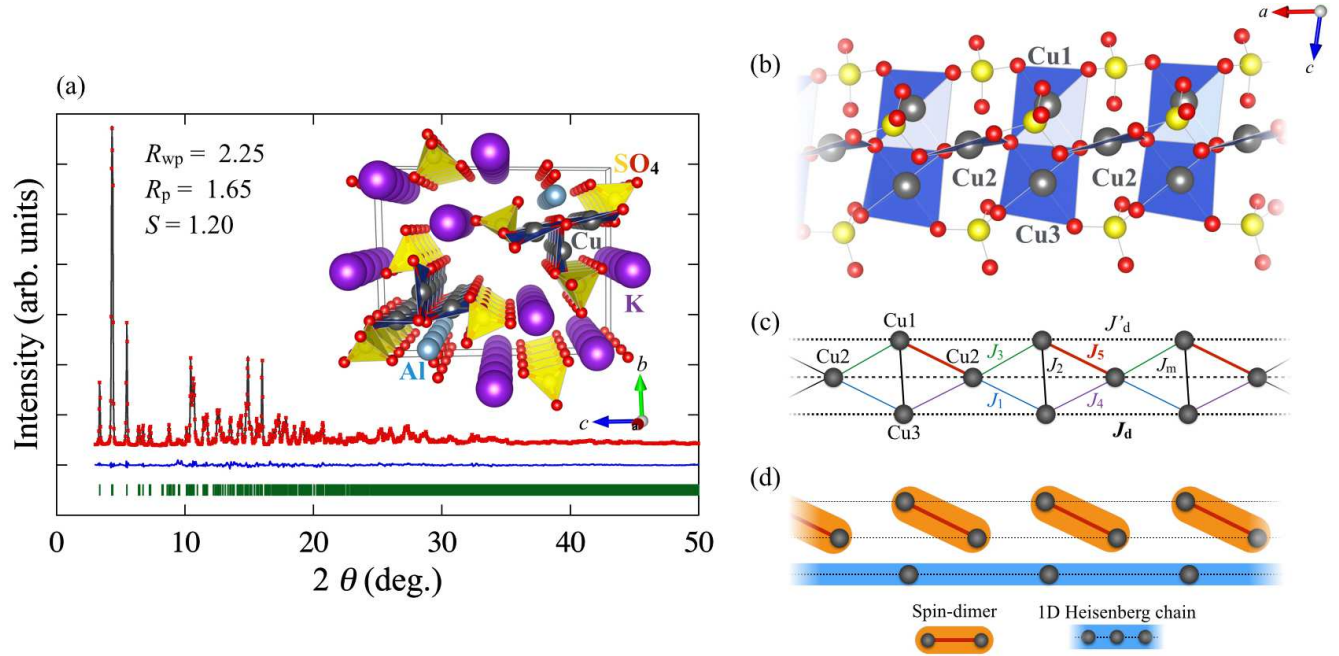}
\caption{(a) Synchrotron XRD intensity pattern (filled red circles) observed for K$_3$Cu$_3$AlO$_2$(SO$_4$)$_4$ at room temperature, the result of Rietveld refinement using the computer program RIETAN-FP~\cite{Fujio} (black solid line), and difference between the calculated and observed intensities (blue solid line). The green vertical bars indicate the position of Bragg reflection peaks. The inset shows the crystal structure of K$_3$Cu$_3$AlO$_2$(SO$_4$)$_4$ featuring a large inter-chain spacing.  
(b) The diamond chain of K$_3$Cu$_3$AlO$_2$(SO$_4$)$_4$, which consists of Cu$^{2+}$ ions (grey spheres) along the $a$-axis with nearby oxygen (red spheres) and sulfate ions (yellow tetrahedral). 
(c) Effective spin model of K$_3$Cu$_3$AlO$_2$(SO$_4$)$_4$ with the nearest-neighbor exchange couplings $J_i$ ($i=1$ to 5), and the next nearest-neighbor exchange couplings of $J_\mathrm{m}$, $J_\mathrm{d}$, and $J'_{\rm d}$. 
(d) Spin configuration of the ground state for K$_3$Cu$_3$AlO$_2$(SO$_4$)$_4$.}
\label{config}
\end{figure}

The Cu-O-Cu angle significantly influences on the value of the exchange interactions, the variation of the angles can give strong bond-dependent exchange interactions~\cite{Mizuno}.
Therefore, the crystal structure refinement of K$_3$Cu$_3$AlO$_2$(SO$_4$)$_4$ is necessary to determine the magnetic exchange interactions.
The space group of the mineral alumoklyuchevskite K$_3$Cu$_3$Al$_{0.64}$Fe$_{0.36}$O$_2$(SO$_4$)$_4$, is $C$2, as reported by Krivovichev $et ~al.$\cite{Krivovichev}.
However, the previous powder XRD data showed that it is not consistent with synthesized K$_3$Cu$_3$AlO$_2$(SO$_4$)$_4$.

The space group and structural parameters for the synthesized material are determined from single crystal XRD and synchrotron powder XRD to be $P$\={1} and $a=4.9338(5)$~\AA, $b=11.923(5)$~ \AA, and $c=14.578(5)$~\AA, with $\alpha=87.309(5)^{\circ}$, $\beta=80.837(3)^{\circ}$, and $\gamma=78.458(6)^{\circ}$, respectively.
As shown by Rietveld refinement results, there is no impurity peaks~[Fig.~\ref{config}(a)]. 
The inset of Fig.~\ref{config}(a) and Fig.~\ref{config}(b) show that K$_3$Cu$_3$AlO$_2$(SO$_4$)$_4$ contains magnetic Cu$^{2+}$ ions in an inequilateral diamond-chain arrangement along the $a$-axis direction.
The nearest-neighbor magnetic couplings, $J_i$ ($i=1$ to 5), are the superexchange interactions through Cu-O-Cu bonds: $J_1$ ($J_4$) through Cu2-O-Cu3 bond with bond angle 101.59(17)$^{\circ}$ (105.38(18)$^{\circ}$), $J_2$ through Cu1-O-Cu3 with two bond angles 96.6(2)$^{\circ}$ and 96.7(2)$^{\circ}$, and $J_3$ ($J_5$) through Cu1-O-Cu2 with  104.51(18)$^{\circ}$ (127.8(2)$^{\circ}$), see Fig.~\ref{config}(c).
In addition, the exchange interactions through the Cu-O-S-O-Cu exchange paths are denoted by $J_{\rm m}$, $J_{\rm d}$, $J'_{\rm d}$ in Fig.~\ref{config}(b).
$J_5$ with the largest angle is expected to be the largest antiferromagnetic (AFM) interaction, while $J_2$ with the smallest angle is considered to be a ferromagnetic (FM) interaction~\cite{Mizuno}.  

The values of the exchange interactions have been obtained by fitting the temperature dependence of the magnetic susceptibility with calculated one~\cite{Morita}: $J_1=J_3=J_4=-30$~K, $J_2=-300$~K, $J_5=510$~K, and $J_{\rm m}=J_{\rm d}=J'_{\rm d}=75$~K.
These values are completely different from those of azurite~\cite{Jeschke}: the $J_2$ bond, where the singlet dimer is located in azurite, is FM, while the singlet dimer and a 1D chain is formed on the $J_5$ bond and $J_{\rm d}$ bond, respectively, as shown in Figs.~\ref{config}(c) and (d).
The alternating dimer-monomer model realized in azurite is, thus, not the case in K$_3$Cu$_3$AlO$_2$(SO$_4$)$_4$.
The TL spin liquid in this compound is formed along the $J_\mathrm{d}$ bond.
We will show below that our experimental results are in excellent agreement with this prediction.%

\subsection*{Spinon continuum and spin-singlet dimer excitations in the inelastic neutron scattering spectra}

\begin{figure}[ht]
\centering
\includegraphics[width=\linewidth]{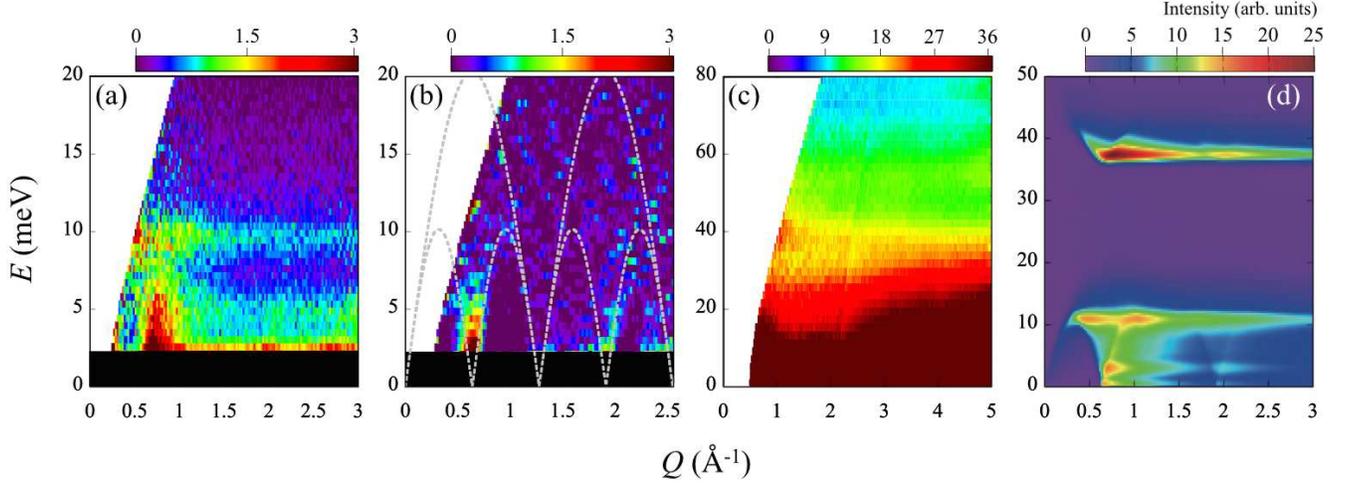}
\caption{INS spectra for K$_3$Cu$_3$AlO$_2$(SO$_4$)$_4$. (a) Magnetic scattering contribution, extracted the 100~K data from the 4~K data corrected for the phonon population factor ($E_i = 45.95$~meV). (b) Single-crystal-like data obtained by applying the conversion method. The superimposed gray dashed lines indicate the lower and upper energy boundaries of the continuum given by $(\pi J_{\rm d}/2)|\sin(Qa)|$ and $\pi J_{\rm d}|\sin(Qa/2)|$~\cite{Lake}, respectively, where $a$ is the lattice parameter in the chain direction. (c) Experimental raw data measured at 4 K with incident neutron energy of 205.8meV. (d) Simulated powder-averaged INS spectrum. }
\label{ins}
\end{figure}

The theoretical study predicted that a gapless low-energy spin excitation and gap excitation are observed by INS experiment, because the spin dimer together with a nearly isolated 1D Heisenberg spin chain characterizes magnetic properties of K$_3$Cu$_3$AlO$_2$(SO$_4$)$_4$~\cite{Morita}.

INS experiments are performed on the powder samples using the high resolution chopper spectrometer HRC at MLF of J-PARC. 
Figure~\ref{ins}(a) shows the magnetic scattering contribution at 4~K, which is obtained by subtracting the phonon contribution from the observed spectrum at 100~K (see Supplementary Information Sec. I).
The strong flat signal is seen at approximately 10~meV, indicating that there is the van Hove singularity of spinon continuum edges at this energy. 
Furthermore, the conversion method developed by Tomiyasu {\it et al}.~\cite{Tomiyasu} is used to obtain single-crystal-like information on magnetic excitations along the chain direction from the powder INS spectrum [Fig.~\ref{ins}(b)].
The spinon continuum edges rise up from the Brillouin zone centers in chain direction $Q=\pi/a=0.64$~\AA$^{-1}$ and $3\pi/a=1.91$~\AA$^{-1}$, which supports the ideal one-dimensionality of this compound. 

Figure~\ref{ins}(c) shows the data measured with $E_i= 205.8$~meV. A signal is observed at around $E=40$~meV and $Q=1.0$~\AA$^{-1}$.
The signal due to magnetic excitations is generally enhanced at low-$Q$ values, whereas phonon excitations are dominant at high-$Q$.
Therefore, we consider that this signal comes from magnetic excitations. 

The dynamical spin structure factor for the proposed model has been calculated by the dynamical density-matrix renormalization group~\cite{Morita}.
In order to compare the calculated spectrum with the powder INS spectra in Figs.~\ref{ins}(a) and \ref{ins}(c), we simulate the powder-averaged INS spectrum including the magnetic form factor of Cu$^{2+}$ by using a conversion technique (see Eq.~(2) of Ref.~13).
Figure~\ref{ins}(d) shows the simulated powder-averaged INS spectrum. The agreement with experimental data in Figs.~\ref{ins}(a) and \ref{ins}(c) is fairly good.
Therefore, we are confident that the proposed exchange interactions and the spin configuration shown in Fig.~\ref{config}(c) are appropriate for this compound.
In other words, the low-energy excitation is characterized by a TL spin liquid.

\subsection*{$\mu$SR evidence for a quantum spin liquid state in K$_3$Cu$_3$AlO$_2$(SO$_4$)$_4$}

\begin{figure}[ht]
\centering
\includegraphics[width=\linewidth]{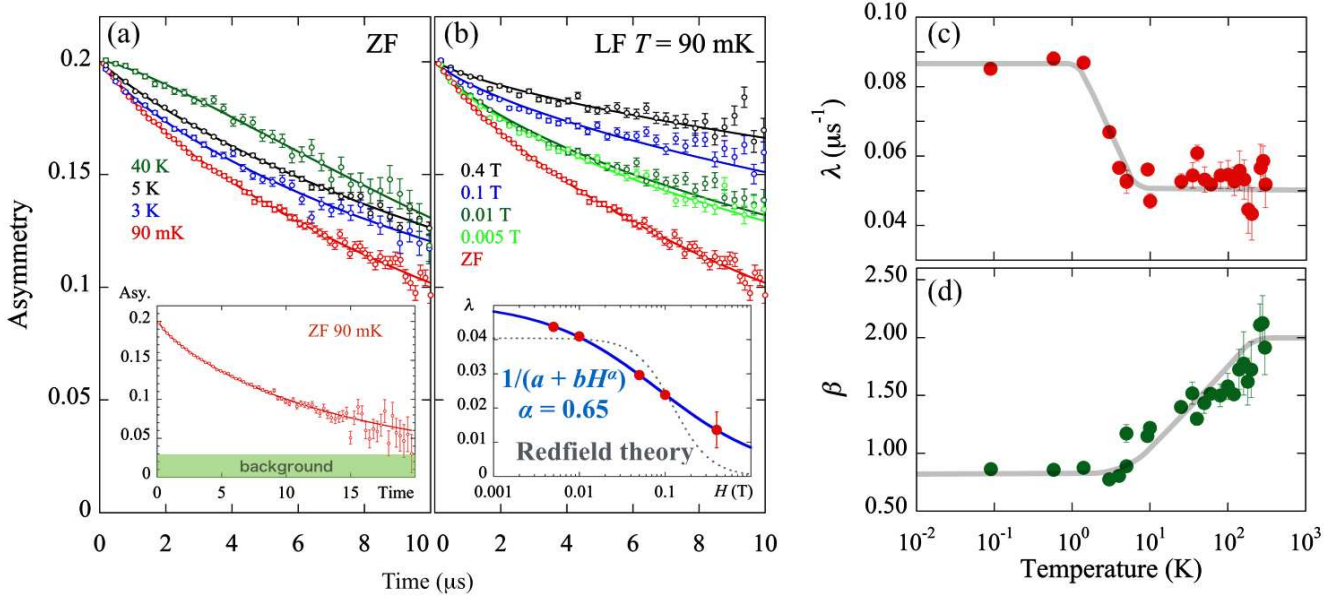}
\caption{(a) ZF-$\mu$SR spectra (using a dilution refrigerator) at representative temperatures (see Supplementary Information Sec. II for the spectra obtained using $^4$He cryostat). The thick lines behind the data points are curves fitted by using the stretched exponential function (see text). The inset shows the ZF-$\mu$SR spectrum measured at 90~mK, which decreases continuously without oscillations up to 20~$\mu$s. (b) $\mu$SR spectra measured at 90~mK under ZF and representative external fields. The inset shows the field dependence of the muon spin relaxation rate $\lambda$. The blue solid line and gray dashed line are curves fitted by using the power law and the Redfield equation (see text). (c) Temperature dependence of $\lambda$. (d) Temperature dependence of the stretching exponent $\beta$. The gray solid lines in (c) and (d) are the guide to eyes.}
\label{msr}
\end{figure}

Quantum spin fluctuations in K$_3$Cu$_3$AlO$_2$(SO$_4$)$_4$ are investigated by using zero-field (ZF) and longitudinal-field (LF) $\mu$SR measurements for a powder sample in the temperature range from 90~mK to 300~K at MLF of J-PARC.

The ZF-$\mu$SR spectra are fitted by the stretched exponential function $a(t)=a_1\exp[-(\lambda t)^{\beta}] + a_{\rm BG}$, where $a_1$ is an intrinsic asymmetry $a_1 = 0.17$, $a_{\rm BG}$ is a constant background $a_{\rm BG} = 0.031$, $\lambda$ is the muon spin relaxation rate, and $\beta$ is the stretching exponent. The spectra at representative temperatures are presented in Fig.~\ref{msr}(a). The combined effect of these multiple nuclear dipole fields leads to a phenomenologically described relaxation function of the stretched exponential. The field distribution and the stretching exponent at high temperatures are approximately given by $\Delta_{\rm nuclear}=\lambda/\gamma_{\mu}=0.6$~G and $\beta\approx 2$, respectively, which are typical for a nuclear dipole field. The ZF spectrum at the lowest temperature $90$~mK decreases continuously without oscillations up to 20~$\mu$s (see the inset of Fig.~\ref{msr}(a)). If this ZF spectrum is due to static magnetism, the internal field should be approximately 1.1~G. However, the relaxation is clearly observed, even in the LF at 0.4~T, which is evidence for the fluctuation of Cu$^{2+}$ electron spins without static ordering down to 90~mK. 

The plateau of the relaxation rate $\lambda$ below 1.5~K and the gradual decrease of $\beta$ are observed, as shown in Figs.~\ref{msr}(c) and \ref{msr}(d), respectively. 
These characteristics have also been observed in other quantum spin-liquid candidates~\cite{Gomilsek,Mendels,Adroja,Kermarrec,Quilliam}. 
The LF spectra at 0.005~T are also fitted by a stretched exponential function with $\beta=0.748$, because there is a complete decoupling of the nuclear component. 
Using the power law represented by $1/(a+ bH^\alpha)$~\cite{Kermarrec,Quilliam} with an unconventional value of $\alpha=0.65$, where $a$ and $b$ are dependent on the fluctuation rate and fluctuating field, we obtain a good fitting to the data, as shown in the inset of Fig.~\ref{msr}(b). 
Incidentally, the $1/(a+ bH^2)$ is a standard case that the $\lambda$ obeys the Redfield equation.
The value obtained for $\alpha$ and the magnitude of the $\lambda$ plateau are close to those for Mg-herbertsmithite~\cite{Kermarrec}. 
All of these $\mu$SR results strongly support the formation of a quantum spin liquid at very low temperature in K$_3$Cu$_3$AlO$_2$(SO$_4$)$_4$.%

\section*{Conclusion}

In summary, a spin-1/2 inequilateral diamond-chain compound K$_3$Cu$_3$AlO$_2$(SO$_4$)$_4$ has been experimentally examined by single-crystal and powder X-ray diffraction, muon spin rotation/relaxation spectroscopy, and inelastic neutron scattering. By comparing the INS experimental data with calculations for a theoretically proposed model, we have confirmed that the compound is described by a composite structure consisting of singlet dimers and a one-dimensional Heisenberg chain, which is different from an alternating dimer-monomer model corresponding to azurite. Since the low-energy excitations are described by the one-dimensional Heisenberg model and there is no three-dimensional long-range order down to 90~mK, K$_3$Cu$_3$AlO$_2$(SO$_4$)$_4$ is regarded as a typical compound that exhibits a TL spin liquid behavior at low temperatures close to the ground state. K$_3$Cu$_3$AlO$_2$(SO$_4$)$_4$ would further contribute to experimental efforts in uncovering exotic properties of the TL spin liquid, such as spinon spin currents~\cite{Hirobe} and ballistic thermal conduction~\cite{Kawamata}.

\section*{Methods}
Single phase polycrystalline K$_3$Cu$_3$AlO$_2$(SO$_4$)$_4$ was synthesized by solid-state reaction in which high-purity K$_2$SO$_4$, CuO, CuSO$_4$, and AlK(SO$_4$)$2$ powders were mixed in a molar ratio of 1:2:1:1, followed by heating at 600$^{\circ}$C for three days and slow cooling in air.
Significant efforts were made to grow single crystals, whereby several tiny single crystals of K$_3$Cu$_3$AlO$_2$(SO$_4$)$_4$ were successfully grown by heating the preliminarily grown powder at temperatures as high as 600$^{\circ}$ in a sealed and evacuated quartz tube for one week with slow cooling.
Single crystal XRD data were collected on a Bruker AXS Smart ApexII ULTRA CCD diffractometer using MoK${\alpha}$ ($\lambda=0.71073$ ~\AA) radiation. Synchrotron powder XRD data were collected using an imaging plate diffractometer installed at BL-8B of the Photon Factory.
An incident synchrotron X-ray energy of 18.0 keV (0.68892~\AA) was selected. 
The INS experiments are performed on the HRC, installed at BL12 beamline at the Materials and Life Science Experimental Facility (MLF) of the Japan Proton Accelerator Research Complex (J-PARC). %
At the HRC, white neutrons are monochromatized by a Fermi chopper synchronized with the production timing of the pulsed neutrons. %
The energy transfer was determined from the time-of-flight of the scattered neutrons detected at position sensitive detectors. %
A 400 Hz Fermi chopper was used to obtain a high neutron flux. A GM-type closed cycle cryostat was used to achieve 100 K and 4 K. The energies of incident neutrons were $E_i$ = 205.8 meV and 45.95 meV (second frame), which yielded an energy resolutions of $E$ = 5 and 1 meV at the elastic position.  %
ZF and LF $\mu$SR experiments are performed using the spin-polarized pulsed surface-muon ($\mu$$^+$) beam at the D1 beamline of the MLF of J-PARC. The spectra were collected in the temperature range from 90 mK to 300 K using a dilution refrigerator and $^4$He cryostat.

\section*{Acknowledgements}

The authors are grateful to S. Takeyama, D. Nakamura, K. Okamoto and T. Goto for helpful discussions. The $\mu$SR and INS experiments were performed at the MLF of J-PARC under a user program (Proposal Nos. 2015A0314, 2016A0130 and 2015A0225). Synchrotron powder XRD measurements were performed with the approval of the Photon Factory Program Advisory Committee (Proposal Nos. 2015P001 and 2016G030). Theoretical study is in part supported by Creation of new functional devices and high-performance materials to support next-generation industries (GCDMSI) to be tackled by using post-K computer and by MEXT HPCI Strategic Programs for Innovative Research (SPIRE) (hp160222, hp170274). This study is partly supported by the Grant-in-Aid for Scientific Research (No. 26287079) and (No. 15H03692) from MEXT, Japan.

\section*{Author contributions statement}

M.F., K.H. and S.M. conceived the study. 
K.M. and T. T. provided theoretical insight and calculated powder-averaged dynamical spin structure factor.
K.T. analyzed the INS data. 
A.K., H.O., M.F. and K.H. performed the $\mu$SR experiments.
S.It, T.Y., S.I., M.F. and K.H. performed the neutron scattering experiments.
M.T. and M.I. collected and analyzed the single-crystal XRD data.
H.S., R.K. and Y.M. helped collect the Synchrotron powder XRD data.
All the authors contributed to the writing of the manuscript. 

\section*{Additional information}
{\bf Supplementary Information} accompanies this paper at http://www.nature.com/srep  \\
\textbf{Competing interests:} The authors declare no competing financial interests. %

\end{document}